\documentclass[english,aps,manuscript,showpacs,superscriptaddress]{revtex4}
\usepackage[T1]{fontenc}
\usepackage[latin9]{inputenc}
\usepackage{verbatim}
\usepackage{amsmath}
\usepackage{amssymb}
\usepackage{graphicx}
\usepackage{esint}
\usepackage{textcomp}
\usepackage{color}

\makeatletter
\@ifundefined{textcolor}{}
{%
 \definecolor{BLACK}{gray}{0}
 \definecolor{WHITE}{gray}{1}
 \definecolor{RED}{rgb}{1,0,0}
 \definecolor{GREEN}{rgb}{0,1,0}
 \definecolor{BLUE}{rgb}{0,0,1}
 \definecolor{CYAN}{cmyk}{1,0,0,0}
 \definecolor{MAGENTA}{cmyk}{0,1,0,0}
 \definecolor{YELLOW}{cmyk}{0,0,1,0}
}

\newif\ifhyper
\hypertrue
\ifhyper       

\usepackage{babel}

\makeatother

\usepackage{babel}
\begin{document}

\title{Global fixed point potential approach to frustrated antiferromagnets}

\author{Shunsuke Yabunaka}

\affiliation{Advanced Science Research Center, Japan Atomic Energy Agency, Tokai, 319-1195, Japan}

\author{Bertrand Delamotte}

\affiliation{Sorbonne Universit\'e, CNRS, Laboratoire de Physique Th\'eorique de la Mati\`ere Condens\'ee, LPTMC, F-75005 Paris, France}

\date{\today}

\begin{abstract}
We revisit the critical behavior of classical frustrated systems using the nonperturbative renormalization group (NPRG) 
equation. Our study is performed within the local potential approximation of this equation
to which is added the flow of the field renormalization. Our flow equations are functional
to avoid possible artifacts coming from the field expansion of the fixed point potential which consists in keeping only  
a limited number of coupling constants.  We explain in detail our numerical implementation, its advantages and the difficulties encountered in the vicinity of $d=2$.  For $N$-component spins, the function $N_c(d)$ separating the regions of first and second order transitions in the $(d,N)$ plane is computed for $d$ between 4 and 2.3. 
Our results confirm what was previously found with cruder approximations of the  NPRG equation
and contradict both the fixed dimension perturbative approach and some of the results
obtained within the conformal bootstrap approach.
\end{abstract}
\maketitle


\section{Introduction}
The critical behavior of antiferromagnetic frustrated systems is still a debated question  more than 
forty years after the first studies of these systems \cite{Delamotte Review,Kawamura Review}. The key difference between 
frustrated and nonfrustrated systems is that the order parameter
is a vector in the nonfrustrated case and a matrix in the other cases. When frustration 
originates from the geometry of the system as in Stacked Triangular Antiferromagnets (STA),
the symmetry of the Hamiltonian is $O(N)\otimes O(2)$ for $N$-component spins and the order
parameter is a rectangular $N\times 2$ matrix \cite{Order parameter symmetry}. Depending on $N$ and the dimension $d$ of space,
the nature of the phase transition changes, being first order for low values of $N$ and dimensions $d$ close to four and second order otherwise. One of the key questions is thus the determination of the 
line $N_c(d)$ separating the first and second order  regions in the $(d,N)$ plane. It turns out that the value of $N_c(d=3)$
is certainly close to 3 and its precise determination is crucial to know whether the transition is
first or second order for the systems realized in nature, that is, for $N=2$ and 3. 
Numerical simulations of several frustrated antiferromagnets such as XY and Heisenberg STA 
have been showing that the transition is first order for these systems \cite{Loison1, Loison2, STA Itakura, Tanh Ngo Diep}. However a recent simulation of Heisenberg STA with a very large lattice size found second order transition corresponding to a focus fixed point (FP) \cite{Nagano Kawamura}.
Depending on the theoretical approach considered, the determination 
of $N_c(d)$ varies much when $d\lesssim 3.3$
 and, as a result, it is not yet settled whether all $O(N)\otimes O(2)$ symmetric systems 
 undergo first order phase transitions in $d=3$ for $N\le3$. The two-dimensional physics
 of the XY and Heisenberg systems is also debated because the relevance of topological defects
 is not yet understood, in particular the possibility that they trigger a phase transition at finite
 temperature \cite{Kawamura Miyashita, Wintel, Stephan, Caffarel, Calabrese focus fixed points 2, Azaria}.
 
The different renormalization group approaches tackling with the problem of the calculation of $N_c(d)$
can be roughly divided into two classes:  the perturbative and the nonperturbative renormalization group (NPRG)
calculations.
The class of perturbative calculations can be again divided into several different subclasses
depending on whether they are performed directly in $d=3$ (at six loops) \cite{Pelissetto,Calabrese focus fixed points,Calabrese 3} or in an $\epsilon$-
or pseudo-$\epsilon$-expansion (respectively at six and five loops)\cite{Calabrese five-loop}. In the latter case, the value of $N_c(d=3)$
is systematically found larger than 3 (of order 6) as it is also the case for the NPRG calculations
that find $N_c(d=3)\simeq 5.1$\cite{Zumbach1, Zumbach2, NPRG field expansion, Delamotte Review, NPRG semi-expansion}. On the contrary, the perturbative calculation performed directly in
$d=3$ at six loops yields a fixed point for $N=2$ and 3 and thus predicts that several $O(N)\otimes O(2)$ symmetric
systems should undergo a second order phase transition. 

Recently, a completely different method based on the conformal bootstrap has been used to study matrix
models in $d=3$ and in particular the $O(N)\otimes O(2)$ frustrated systems \cite{Nakayama Ohtsuki-1, Nakayama Ohtsuki-2, Stergiou, Rychkov}. This approach has the advantage of being unbiased by convergence 
problems since it is not based on series expansions, contrary to RG methods and, when applied to
the ferromagnetic $O(N)$ models, it leads to a very accurate determination of the critical
exponents, at least when it is truncated at large orders \cite{Bootstrap Ising, Bootstrap Ising-2}.   However for the frustrated systems, the situation is less clear. An early conformal bootstrap study conjectured the existence of a critical FP for $N=2$ and $3$ in $d = 3$, based on kinks of rigorous bounds on scaling dimensions. The predicted critical exponents are in good agreement with those for the focus type of FPs found with the perturbative fixed dimensional approach \cite{Nakayama Ohtsuki-1, Nakayama Ohtsuki-2}. However, by construction, the conformal bootstrap cannot find a fixed point with imaginary exponents because this contradicts reflection positivity. It is pointed out in Section 4 of  \cite{Rychkov} that interpreting a kink as an indication of the existence of a critical FP may not be valid in general.  In a recent refined conformal bootstrap study, rigorous bounds are derived to isolate allowed regions in the space of scaling dimensions and it is shown that a lower bound for $N_c(d)$ for systems satisfying reflection positivity is $N_c(d=3)>3.78$ \cite{Rychkov}.

As for the NPRG approach, that we re-examine here, the situation is the following. Either 
the conclusions drawn from its results are correct and then both the fixed dimension perturbative RG approach  and some of the conformal bootstrap studies \cite{Nakayama Ohtsuki-1,Nakayama Ohtsuki-2} are wrong or, conversely, it is wrong (together with the 
$\epsilon$-expansion approaches) and this implies that
the approximations used so far are too drastic to reproduce the correct physics. In both cases, something very unusual
is at work because the methodologies that have been used in these studies 
lead in many cases to correct and accurate results, see for instance \cite{NPRG}.

The NPRG is based on an exact RG equation that requires approximations to be solved. The approximations used so far
to tackle with frustrated systems consists in performing a derivative expansion \cite{Berges} and a field expansion
of the Gibbs free energy \cite{NPRG field expansion, Delamotte Review, NPRG semi-expansion}. The rationale behind this choice is  (i) that the critical behavior of thermodynamic quantities
such as the specific heat or the susceptibility for instance are dominated by long wavelength fluctuations
which justifies expanding the correlation functions in their momenta (derivative expansion) and (ii) 
that the impact of the $n$-point functions with $n$ large on the RG flow
of the zero or two-point functions should be small (field-expansion). It is the aim of this article to
eliminate one source of inaccuracy of the NPRG approach, the field expansion, which is known to be inaccurate
in low dimensions even for simple models such as the ferromagnetic $O(N)$ models \cite{Delamotte Review}. The price to pay to get
rid of this approximation is to work functionally, that is, to follow the RG flow of functions of
the fields instead of a limited number of coupling constants. In the case of nonfrustrated systems,
this is relatively simple since the $O(N)$ symmetry implies that all functions involved in 
the RG flows depend on the fields
only through the unique $O(N)$-invariant: $\rho={\vec\phi}^{\,2}$. For frustrated systems,
there exists two $O(N)\otimes O(2)$ invariants and the resulting flow equations are partial differential
equations that are rather involved to solve numerically. We show in this article how to simplify the numerical problem
and point out why the  numerical difficulties are so great in low dimensions that our method does no  longer work when approaching $d=2$. We provide the results thus obtained for the curve $N_c(d)$  between $d=4$ and $d=2.3$. 

We note that in previous NPRG studies involving a field expansion of the potential, it was not possible to study the curve $N_c(d)$ below $d=3$ in a reliable manner. One of the expected scenarios to allow for the existence of a FP that drives a second order transition in $d=3$ and $N=3$ is that $N_c(d)$ does not decrease monotonously but has a turn around point around $d=3$, see for instance \cite{Rychkov}. In this scenario, the curve $N_c(d)$ is continuous but has a S-shape, that is, the first-order region is re-entrant around $d=3$ and the function $N_c(d)$ is  multi-valued in $d=3$: when $N$ is large, the transition is second-order, becomes first-order for smaller values of $N$, typically $N<6$, and becomes second-order again for even smaller values of $N$, in particular $N=2$ and 3. Therefore, it is interesting for physics in $d=3$ to study the part of the curve $N_c(d)$ below $d=3$ to find out whether or not such a turn around point exists or the curve is monotonous down to $d=2$. Our results confirm what was previously found within a NPRG approximation involving a  field expansion of the potential
and the $\epsilon$-approaches and thus contradict both the fixed-dimension perturbative approach and the results
obtained with the conformal bootstrap in \cite{Nakayama Ohtsuki-1,Nakayama Ohtsuki-2}.

\section{The Model}

As the archetype of frustrated spin systems, we employ the Stacked
Triangular Antiferromagnets (STA). This system is composed of two-dimensional
triangular lattices that are piled-up in the third direction. At each
lattice site $i$, is defined a  $N$-component
vector $\mathbf{S}_{i}$ of modulus 1.  The Hamiltonian of this system
is given by

\begin{equation}
H=J \sum_{\left\langle ij\right\rangle }\mathbf{S}_{i}\cdot\mathbf{S}_{j}.
\end{equation}
The sum $\left\langle ij\right\rangle $ runs on all pairs of nearest
neighbor spins and $J>0$.

The long distance effective theory for the STA has been derived by
Yosefin and Domany\cite{Order parameter symmetry}. The order parameter consists of
the $N\times2$ matrix $\Phi=\left(\mathbf{\boldsymbol{\phi}}_{1},\mathbf{\boldsymbol{\phi}}_{2}\right)$
that satisfies 
\begin{equation}
\boldsymbol{\phi}_{i}\cdot\boldsymbol{\phi}_{j}=\delta_{ij}
\end{equation}
for $i,j=1,2$. Then, the effective Hamiltonian in the continuum is
given by 
\begin{equation}
H=\int d^{d}\mathbf{x}\left(\frac{1}{2}\left[\left(\partial\boldsymbol{\phi}_{1}\right)^{2}+
\left(\partial\boldsymbol{\phi}_{2}\right)^{2}\right]\right).
\end{equation}

The constraint $\boldsymbol{\phi}_{i}\cdot\boldsymbol{\phi}_{j}=\delta_{ij}$
for $i,j=1,2$ can be replaced by a soft potential 
$U\left(\mathbf{\boldsymbol{\phi}}_{1},\mathbf{\boldsymbol{\phi}}_{2}\right)$
whose minima are given by $\boldsymbol{\phi}_{i}\cdot\boldsymbol{\phi}_{j}\propto\delta_{ij}$
and the Ginzburg-Landau-Wilson Hamiltonian for STA reads 
\begin{equation}
H=\int d^{d}\mathbf{x}\left(\frac{1}{2}\left[\left(\partial\boldsymbol{\phi}_{1}\right)^{2}+
\left(\partial\boldsymbol{\phi}_{2}\right)^{2}\right]+U\left(\mathbf{\boldsymbol{\phi}}_{1},\mathbf{\boldsymbol{\phi}}_{2}\right)\right).\label{eq:effective hamiltonian}
\end{equation}
Instead of $\mathbf{\boldsymbol{\phi}}_{i}$, it is convenient to
work with the invariants of the O$(N)\times$O(2) group that can be
chosen as: 
\begin{equation}
\begin{array}{ll}
\rho= & \mathrm{Tr}\left(^{t}\Phi\Phi\right)=\boldsymbol{\phi}_{1}^2+ \boldsymbol{\phi}_{2}^2,\\
\tau= & \frac{1}{2}\mathrm{Tr}\left(^{t}\Phi\Phi-\rho/2\right)^{2}=
\frac{1}{4}\left(\boldsymbol{\phi}_{1}^2- \boldsymbol{\phi}_{2}^2\right)^2
+\left(\boldsymbol{\phi}_{1}. \boldsymbol{\phi}_{2}\right)^2.
\end{array}
\end{equation}
With this choice, the ground state configuration corresponds to $\rho={\rm const.}$
and $\tau=0$.  Up to the fourth order $U\left(\rho,\tau\right)$
can be written as  
\begin{equation}
U\left(\rho,\tau\right)=\frac{\lambda}{2}\left(\rho-\kappa\right)^{2}+\mu\tau,
\label{pot}
\end{equation}
where $\lambda$ and $\mu$ are positive coupling constants. A typical
ground state in terms of $\Phi$  is
given by $\Phi_{\alpha,i}=\sqrt{\kappa/2}\delta_{\alpha,i}$, that is:
\begin{equation}
\Phi_{\rm min}\equiv\left(\begin{array}{cc}
\sqrt{\frac\kappa 2} & 0\\
0 & \sqrt{\frac\kappa 2}\\
\vdots & \vdots\\
0 & 0
\end{array}\right).
\label{eq:min}
\end{equation}

\section{The nonperturbative renormalization group equation}

The NPRG method is based on Wilson's idea of integrating statistical
fluctuations step by step. In this paper, we employ the effective
average action method as an implementation of the NPRG in continuum
space \cite{Wetterich1,Ellwanger,Morris, Wetterich2}.

 The first step is to
introduce a $k$-dependent partition function $\mathcal{Z}_{k}$ in
the presence of sources: 
\begin{equation}
\mathcal{Z}_{k}\left[\boldsymbol{J}_{i}\right]=
\int\mathcal{D}\boldsymbol{\phi}_{i}\exp\left(-H[\boldsymbol{\phi}_{i}]-
\Delta H_{k}[\boldsymbol{\phi}_{i}]+\boldsymbol{J}_{i}\cdot\boldsymbol{\phi}_{i}\right),
\end{equation}
{where $\mathbf{J}_{i}\cdot\boldsymbol{\phi}_{i}=
\sum_{i=1}^{2}\int_{x}\mathbf{J}_{i}\left(\mathbf{x}\right)\cdot\boldsymbol{\phi}_{i}\left(\mathbf{x}\right),$
and $\Delta H_{k}=1/2\sum_{i=1}^{2}\boldsymbol{\phi}_{i}(x) R_{k}(x-y)\boldsymbol{\phi}_{i}(y)$.}
 The idea underlying the effective average action is that
in $\mathcal{Z}_{k}$ only the fluctuations of large wave-numbers
(the rapid modes) compared to $k$ are integrated over while  the others (the slow modes)
are frozen by the $\Delta H_{k}$ term. As $k$ is decreased, more and more modes are integrated
until they are all when $k=0$. The function $R_{k}({q}^{2})$, which is the
Fourier transform of $R_{k}({x})$, plays the role of separating rapid
and slow modes: It almost vanishes for $\vert q\vert>k$ so that the
rapid modes are summed over and is large (of order $k^{2}$) below
$k$ so that the fluctuations of the slow modes are frozen. We define
as usual $W_{k}[\boldsymbol{J}_{i}]=\ln\mathcal{Z}_{k}[\boldsymbol{J}_{i}]$.
 Thus, the order parameter $\boldsymbol{\varphi}_{j}\left(\mathbf{x}\right)$
at scale $k$ is defined by 
\begin{equation}
{\boldsymbol{\varphi}_{i}\left(\mathbf{x}\right)}=\left\langle \boldsymbol{\phi}_{i}
\left(\mathbf{x}\right)\right\rangle =
\frac{\delta W_{k}\left[\boldsymbol{J}_{i}\right]}{\delta\boldsymbol{J}_{i}\left(\mathbf{x}\right)}\, .
\label{eq:def=00003D00003D0000A5phi}
\end{equation}
The running effective average action $\Gamma_{k}\left[\mathbf{\boldsymbol{\varphi}}_{i}\right]$
is defined  as the (modified) Legendre transform of $W_{k}$:
\begin{equation}
\Gamma_{k}\left[\mathbf{\boldsymbol{\varphi}}_{i}\right]=-W_{k}\left[\boldsymbol{J}_{i}\right]+\mathbf{J}_{i}\cdot\boldsymbol{\varphi}_{i}-\Delta H_{k}\left[\mathbf{\boldsymbol{\varphi}}_{i}\right]
\end{equation}
where $\boldsymbol{J}_{i}$ is defined such that Eq. (\ref{eq:def=00003D00003D0000A5phi})
holds for fixed $\mathbf{\boldsymbol{\varphi}}_{i}$. From this definition
one can show that 
\begin{equation}
\begin{cases}
\Gamma_{k=\Lambda}\simeq H\\
\Gamma_{k=0}=\Gamma
\end{cases},\label{gammakboundaries}
\end{equation}
where the cutoff $\Lambda$ is the inverse of the lattice spacing
$a$. Equations (\ref{gammakboundaries}) imply that $\Gamma_{k}$
interpolates between the Hamiltonian of the system when no fluctuation
has been summed over, that is, when $k=\Lambda$, and the Gibbs free
energy $\Gamma$ when they have all been integrated, that is, when
$k=0$. We define the variable $t$, called ``RG time'', by $t=\ln\left(k/\Lambda\right)$.
The exact flow equation for $\Gamma_{k}$ reads \cite{Wetterich1,Wetterich2}:
\begin{equation}
\partial_{t}\Gamma_{k}[\mathbf{\boldsymbol{\varphi}}_{i}]=
\frac{1}{2}\mathrm{Tr}\int_{x,y}\partial_{t}R_{k}(x-y)\left(\frac{\delta^{2}\Gamma_{k}
\left[\mathbf{\boldsymbol{\varphi}}_{i}\right]}{\delta\varphi_{i}^{\alpha}
\left(\mathbf{x}\right)\delta\varphi_{i'}^{\alpha'}\left(\mathbf{y}\right)}+R_{k}
\left(\mathbf{x-y}\right)\delta_{i,i'}\delta_{\alpha,\alpha'}\right)^{-1},
\label{floweq}
\end{equation}
for $\alpha,\alpha'=1,2,\cdots N$ and $i,i'=1,2$.

\section{Truncations of the NPRG equation}

It is generally not possible to solve exactly the above flow equation (\ref{floweq})
and  approximations are required in practice. In this paper,
we employ the  approximation of lowest level in the derivative expansion
dubbed the local potential approximation (LPA) and some of its refinements.

Within the LPA, $\Gamma_{k}$ is approximated by a series expansion
in the gradient of the field, truncated at its lowest non trivial
order: 
\begin{equation}
\Gamma_{k}^{\rm LPA}\left[\mathbf{\boldsymbol{\varphi}}_{i}\right]=
\int d^{d}\mathbf{x}\left(\frac{1}{2}\left[\left(\partial\boldsymbol{\varphi}_{1}\right)^{2}+
\left(\partial\boldsymbol{\varphi}_{2}\right)^{2}\right]+U_{k}\left(\rho,\tau\right)\right).
\end{equation}
Only a potential term $U_{k}\left(\rho,\tau\right)$ is thus retained in this approximation
which is accurate as long as the impact of the renormalization of the derivative terms on the flow of the potential
is small. This is most probably the case when the anomalous dimension is small and $d>2$. The next level
of approximation consists in including in the approximation a running field renormalization 
$Z_k$ and a coupling constant $\omega_k$, which affects the spectrum of a Goldstone mode around the minimum of the potential and is known to be important for the physics near $d=2$ \cite{Delamotte Review}.
\begin{eqnarray}
\Gamma_k=\int_x \Big\{U_k(\rho,\tau)+\frac{1}{2}
Z_k\Big(\big(\partial \boldsymbol{\varphi}_1\big)^2+ \big(\partial \boldsymbol{\varphi}_2\big)^2\Big)
+ \frac{1}{4} \omega_k \big(\boldsymbol{\varphi}_1\cdot\partial \boldsymbol{\varphi}_2- \boldsymbol{\varphi}_2\cdot\partial\boldsymbol{\varphi}_1\big)^2 \Big\} \ .
\label{action_generale2}
\end{eqnarray}
This approximation  has been used in \cite{tissier00b,tissier01, tissier03,  NPRG field expansion, Delamotte Review, NPRG semi-expansion}  
where  the function $U_k(\rho,\tau)$ was further expanded in powers of the invariants $\rho$ and $\tau$. 
This is what we  improve here to avoid any artifact coming from this field truncation.
This approximation, that we call LPA' with $\omega_k$, yields the one-loop result obtained  
within the $\epsilon$-expansion in   $d=4-\epsilon$ and also
the one-loop result of the $\epsilon=d-2$ expansion of the nonlinear sigma model \cite{Delamotte Review}.  To examine the impact of including the coupling constant $\omega_k$, we also make calculations setting $\omega_k=0$ in Eq. (\ref{action_generale2}): this approximation is called the LPA'.

The $k$-dependent effective potential $U_{k}\left(\rho,\tau\right)$
is defined by
\begin{equation}
\Omega U_{k}\left(\rho,\tau\right)=\Gamma_{k}\left[\mathbf{\boldsymbol{\varphi}}_{i}\right]
\end{equation}
where $\boldsymbol{\varphi}_{i},i=1,2$ are constant fields and $\Omega$
is the volume of the system. The running field renormalization $Z_{k}$ 
is set to one in LPA: $Z_{k}^{{\rm LPA}}=1$, which leads
to a vanishing anomalous dimension: $\eta=0$. In LPA' (or LPA' with $\omega_k$) calculations,
the anomalous dimension $\eta$ is obtained from the flow of $Z_{k}$
since it can be shown that at criticality: 
\begin{equation}
Z_{k\rightarrow0}\sim\left(\frac{k}{\Lambda}\right)^{-\eta}.
\end{equation}
The flows of $U_k$, $Z_k$ and $\omega_k$
have been derived in \cite{NPRG field expansion, Delamotte Review, NPRG semi-expansion} and we give the  expression in terms of $U_{k}\left(\rho,\tau\right)$ after some simplification in 
Appendix A.
These flows are rather complicated and their numerical integration suffers from all the inherent
difficulties of solving nonlinear partial differential equations. 

The first difficulty comes from the choice of variables. 
It is tempting to work with the
invariants $\rho$ and $\tau$ defined above because the symmetry of the problem is encoded 
in the very definition of the variables and any smooth function of these variables corresponds to 
a function that has the right symmetry. However, $\rho$ and $\tau$ satisfy $\frac{1}{4}\rho^{2}\geq\tau\geq0$
and it is not easy to deal with this constraint numerically because the domain where the
variables $\rho$ and $\tau$ live is nontrivial. Thus, we define another
set of variables $\psi_{i}$ which is numerically more convenient.
For any $\boldsymbol{\varphi}_{1}$ and $\boldsymbol{\varphi}_{2}$,
it can be proven that there exists $O_{1}\in O\left(N\right)$ and
$O_{2}\in O\left(2\right)$ such that the matrix $M\equiv O_{1}\Psi O_{2}$, with the $N\times2$ matrix $\Psi$  defined as
$\Psi=\left(\mathbf{\boldsymbol{\varphi}}_{1},\mathbf{\boldsymbol{\varphi}}_{2}\right)$,
becomes ``diagonal'', namely, 
\begin{equation}
M\equiv\left(\begin{array}{cc}
\psi_{1} & 0\\
0 & \psi_{2}\\
\vdots & \vdots\\
0 & 0
\end{array}\right).
\end{equation}
Because of the  $O\left(N\right)\times O\left(2\right)$ symmetry of the model, we conclude
that $U_{k}\left(\Psi\right)=U_{k}\left(M\right)$. This fact shows
that we can parametrize the order parameter space using $\psi_{1}$
and $\psi_{2}$, instead of $\boldsymbol{\varphi}_{1}$ and $\boldsymbol{\varphi}_{2}$.
The $O\left(N\right)\times O\left(2\right)$ invariants $\rho$ and
$\tau$ are expressed in terms of $\psi_{1}$ and $\psi_{2}$ as 
\begin{equation}
\begin{array}{ll}
\rho & =\psi_{1}^{2}+\psi_{2}^{2}\\
\tau & =\frac{1}{4}\left(\psi_{1}^{2}-\psi_{2}^{2}\right)^{2}.
\end{array}\label{eq:def_rho_tau}
\end{equation}
 From the definitions (\ref{eq:def_rho_tau}), we find that the symmetries of the original problem imply:  
\begin{equation}
U_{k}\left(\psi_{1},\psi_{2}\right)=U_{k}\left(-\psi_{1},\psi_{2}\right)=
U_{k}\left(\psi_{1},-\psi_{2}\right)=U_{k}\left(\psi_{2},\psi_{1}\right).\label{constraint}
\end{equation}
 Thus, to solve the flow equations, it is sufficient to consider the region $\psi_{2}\geq\psi_{1}\geq0$.
This triangular domain is much more convenient from a numerical
point of view than the parabolic domain $\frac{1}{4}\rho^{2}\geq\tau\geq0$
for the invariants $\rho$ and $\tau$.

When the transition is of second order, the $k$-dependent effective action is attracted at criticality towards
the fixed point solution of the NPRG flow equation  once it is expressed
in terms of the dimensionless renormalized fields $\tilde{\psi}_{i}$
and a dimensionless local potential $\tilde{U}_{k}(\tilde{\psi}_{i})$.
We thus define the dimensionless and renormalized quantities: 
\begin{equation}\label{tilde-varibales}
\begin{array}{l}
\tilde{\psi}_{i}=\left(Z_{k}k^{2-d}\right)^{1/2}\psi_{i}\\
\tilde{U}_{k}(\tilde{\psi}_{i})=k^{-d}U_{k}\left(\psi_{i}\right).
\end{array}
\end{equation}
The flow equation for $\tilde{U}_{k}$ is given by Eq. (\ref{eq:flowU})
in Appendix A. The critical exponent $\nu$ of the correlation length
is obtained from the relevant eigenvalue of the linearized flow around
the fixed point solution and $\eta$ from the flow of $Z_{k}$. The
other leading critical exponents can be deduced from these ones by scaling
relations.

Throughout this paper we employ the following $R_{k}\left(\mathbf{q}^{2}\right)$: 
\begin{equation}
R_{k}\left(\mathbf{q}^{2}\right)=\beta Z_{k} k^2\left(1-\frac{\mathbf{q}^{2}}{k^2}\right)^{\alpha}\Theta\left(k^{2}-\mathbf{q}^{2}\right),
\label{reglitim}\end{equation}
where we have introduced the parameters $\alpha\ge1$ and $\beta>0$, and $Z_{k}$ is defined as 
\begin{equation}
Z_{k}=\left(\frac{\partial}{\partial p^{2}}\left(\frac{\delta^{2}\Gamma_{k}}{\delta\varphi_{1}^{3}\left(\mathbf{p}\right)\delta\varphi_{1}^{3}\left(-\mathbf{p}\right)}/\left(2\pi\right)^{d}\delta\left(\mathbf{0}\right)\right)\right)_{\mathbf{p}=0,{\rm min}},
\end{equation}
where the field values are set to the minimum of $U_{k}$ given by Eq. (\ref{eq:min}). The regulator function $R_{k}\left(\mathbf{q}^{2}\right)$ is useful for analytical treatments when $\alpha=\beta=1$ \cite{Litim} and is probably optimal at  LPA \cite{Canet-LPA}. \textcolor{black}{We note that this family of regulators yields very accurate values of the critical exponents for $O(N)$ models at the fourth order of derivative expansion, after optimization  based on PMS on $\alpha$ and $\beta$, as shown \cite{Canet-Ising, Balog, DePolsi}), for example. Therefore we expect that Eq. (\ref{reglitim}) is also a good choice of regulator for $O(N)\times O(2)$ models.} Here the Fourier transform $\phi_{1}^{3}\left(\mathbf{p}\right)$
is defined as $\phi_{1}^{3}\left(\mathbf{p}\right)=\int d^{d}\mathbf{x}\phi_{1}^{3}\left(\mathbf{x}\right)\exp\left(-i\mathbf{x}\cdot\mathbf{q}\right).$ 
\textcolor{black}{The value of any physical quantity computed with a given regulator is independent of this regulator  if the derivative expansion is not truncated to a finite order. However we  do truncate it by employing the LPA, LPA' and LPA' with $\omega_k$, which are different approximations of the derivative expansion of the wavenumber dependent effective action $\Gamma_k$. Since the results at each level of the approximation are not exact, they  depend on the choice of regulator,} \textcolor{black}{which explains why we perform the PMS optimization.}

The dimensionless $O\left(N\right)\times O\left(2\right)$ invariants $\tilde{\rho}$
and $\tilde{\tau}$ are defined by $\tilde\rho=Z_k k^{2-d}\rho$, $\tilde\tau=Z^2_k
k^{2(2-d)}\tau$, and the potential and couplings by $\widetilde{U}_k(\tilde\rho,\tilde\tau)=k^{-d}{U}_k(\rho,\tau)$,  
$\widetilde \omega_k=Z^{-2}_k k^{d-2} \omega_k$, 
$y=q^2/k^2$,  $R_k(q^2)=Z_k k^2 y r(y)$. Notice that as said above,
$Z_k$ does not reach a fixed point but  $\eta_k$, defined by $\eta_k=-d\log Z_k/d\log k$, does: 
$\eta_{k\to0}\to \eta$ at criticality with $\eta$ the anomalous dimension of the fields. The flow of $\tilde{\omega}_k$ is also evaluated at the minimum of the potential.

\section{Numerical results}

As said above, we use the variables $\tilde\psi_i$ defined in Eq.~\eqref{tilde-varibales} to integrate the FP equations. These equations are discretized on a regular lattice and the domain corresponding to our numerical grid is a triangle in the  $(\tilde\psi_1,\tilde\psi_2)$ plane of linear size $\tilde{\psi}_{\rm max}$ and mesh $\Delta \psi$. We have checked the convergence of all the numerical results presented below by varying these parameters.

\subsection{The numerical method used to calculate the line $N_c(d)$}

The line $N_c(d)$ separates in the $(d,N)$ plane the region where the phase transition 
is of second order and the region where it is of first order. When the transition is of second order, a once-unstable FP called $C_+$ is associated with it. Two other nontrivial FPs  are also found: the O$(2N)$ FP and a tricritical FP called $C_-$. The line $N_c(d)$ is the locus in the $(d,N)$ plane where $C_+$ disappears by colliding with $C_-$: below this line, the RG flow \textcolor{black}{was found to} show a runaway which is the hallmark of a  first-order transition \textcolor{black}{\footnote{ When a  field expansion of the potential is performed, it has been confirmed that the corremlation length remains finite around $d=3$ by integrating the renormalization group flow as shown in \cite{Delamotte Review}. We think that it is interesting to study  runaway flows with the current treatment of the full functional dependence of $U_k(\rho,\tau)$ in future studies.}}.

There are two possibilities to determine $N_c(d)$. Either we decrease $N$ at fixed $d$ and look for
the value of $N$ where $C_{+}$ is no longer found and then repeat the same procedure by decreasing $d$.
Or we compute the smallest eigenvalue of the flow around the fixed point $C_{+}$ corresponding
to an irrelevant direction and look for the value of $N$ where it vanishes. This eigenvalue is 
a measure of the speed of the flow on the RG trajectory joining $C_{+}$ and $C_{-}$ and
this speed goes to 0 when the fixed points collapse. We have used both methods and, for the precision required in the present work, they are both satisfactory. 

\subsection{Calculation of $N_c(d=3)$}
\begin{figure}[t!]
	\centering 
    \includegraphics[width=0.6\textwidth]{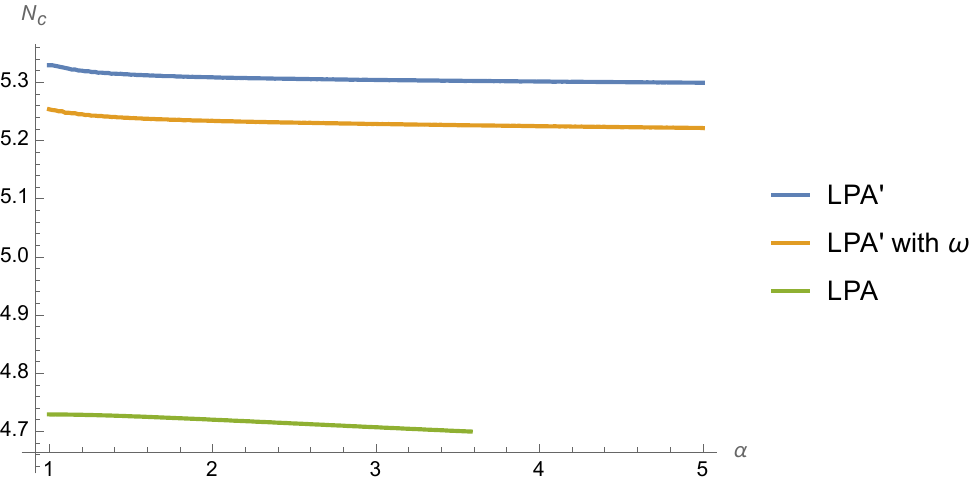}
    \caption{The curve $N_c(d=3, \beta=1, \alpha)$ at LPA, LPA' and LPA' with $\omega_k$. Here we fix the parameter $\beta=1$  and vary the parameter $\alpha$ in the regulator $R_k(q^2)$ defined in Eq.~\eqref{reglitim}. }
    \label{Ncd=3}
\end{figure}

We show $N_c(d=3, \beta=1, \alpha)$ at LPA, LPA' and LPA' with $\omega_k$ in Fig \ref{Ncd=3}. This quantity  does not have an extremum as a function of $\alpha$. When such a quantity has an extremum as a function of a tunable and unphysical parameter such as $\alpha$, this extremum is interpreted as the optimal value of this quantity because it corresponds to the region of smallest dependence upon the parameter (Principle of Minimal Sensitivity). When applied to $O(N)$ models, this principle allows  accurate determination of the critical exponents \cite{Canet-Ising,Balog,DePolsi}. As can be seen on Fig.~\ref{Ncd=3}, there is no extremum but the magnitude $\Delta N_c$ of the variation of $N_c$ is very small, smaller than 0.1, when $\alpha$ is varied on a large range of values, typically between 1 and 5. Consequently, the uncertainty associated with the choice of regulator seems much smaller than that associated with the truncation of the DE at either LPA or LPA' or LPA' with $\omega_k$, considering the variations of this quantity from one approximation scheme to another. We therefore retain the values: $N_c(d=3)\simeq 4.7$, $5.3$ and $5.2$ for respectively the LPA, LPA' and LPA' with $\omega_k$. We note that Zumbach in \cite{Zumbach1} already obtained $N_c(d=3)\simeq 4.7$ at LPA implemented on the Polchinski equation without any field expansion. These values of $N_c(d=3)$ are almost  compatible  with $N_c^{\rm semi}(d=3)\simeq 4.68(2)$, $5.24(2)$ obtained from a ``semi-expansion''  \cite{NPRG semi-expansion} within LPA and LPA' with $\omega_k$, respectively. In the semi-expansion approximation,  the potential is treated functionally in the $\rho$-direction  but expanded  in the $\tau$-direction. This expansion {is found to converge nicely at small orders in $\tau$} for $d\ge3$. In \cite{NPRG semi-expansion}, another regulator called the exponential regulator was employed, instead of Eq. (\ref{reglitim}) used in the present study:  
\begin{equation}
R_{k}^{\rm exp}\left(\mathbf{q}^{2}\right)=Z_{k} \alpha' \mathbf{q}^{2}/(e^{\mathbf{q}^{2}/k^2}-1).
\label{regexp}\end{equation}
For this regulator, the semi-expansion is expected to almost converge in $d=3$ within LPA and LPA' with $\omega_k$. 
Therefore, the fact that the two studies give almost the same values of $N_c(d)$ suggests that the dependence of $N_c(d)$ on the choice of the regulator is very small and $N_c(d)$ is rather precisely determined within each approximation (LPA, LPA' or LPA' with $\omega_k$), at least for $d\ge3$. The remaining source of error is the field dependence of the second order derivative terms, which is not fully taken into account in the current approximations, and the  higher order derivative terms in $\Gamma_k$. Therefore, the discrepancy on $N_c(d)$ between NPRG and the $\epsilon-$expansion might be improved with full calculation at second order of derivative expansion on the side of NPRG, as a first step.

\subsection{Critical exponents $\nu$ and $\eta$ for $N=6$ and $d=3$}

\begin{figure}[t!]
	\centering 
    \includegraphics[width=0.6\textwidth]{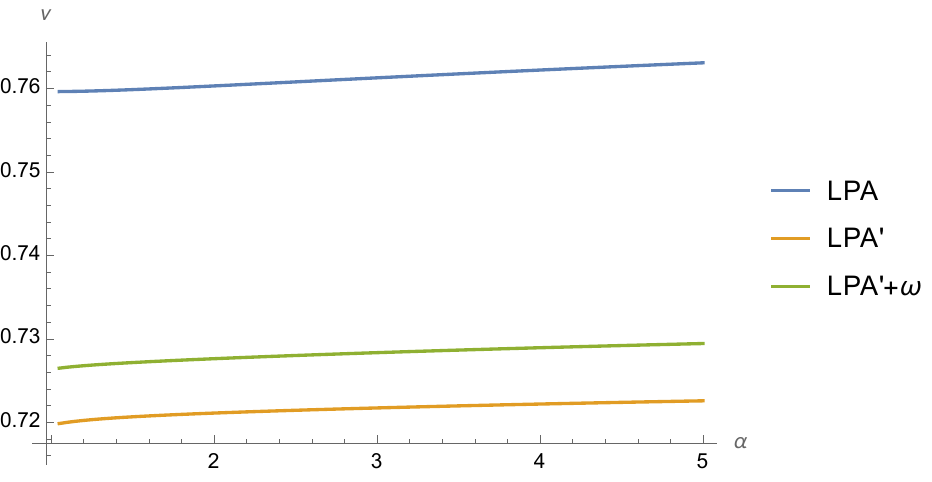}
    \includegraphics[width=0.6\textwidth]{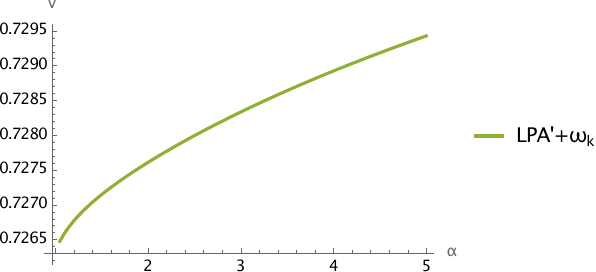}
    \caption{The critical exponent $\nu(d=3, N=6, \beta=1, \alpha)$ at LPA, LPA' and LPA' with $\omega_k$. Here we fix the parameter $\beta=1$  and vary the parameter $\alpha$ in the regulator $R_k(q^2)$. We do not have PMS.}
    \label{nu-n=6}
\end{figure}

\begin{figure}[t!]
	\centering 
    \includegraphics[width=0.6\textwidth]{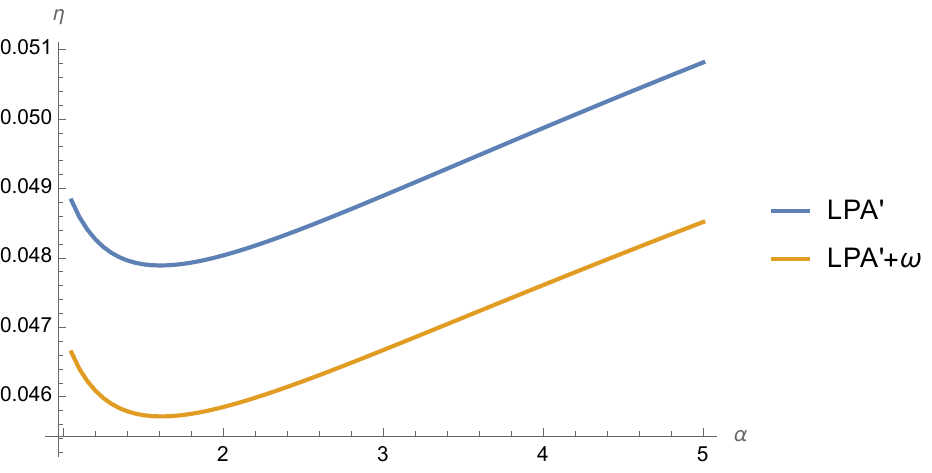}
    \caption{The critical exponent $\eta(d=3, N=6, \beta=1, \alpha)$ at LPA' and LPA' with $\omega_k$. Here we fix the parameter $\beta=1$  and vary the parameter $\alpha$ in the regulator $R_k(q^2)$. We have PMS around $\alpha=1.6$.}
    \label{etad=3}
\end{figure}

We show the critical exponent $\nu(d=3, N=6, \beta=1, \alpha)$ at LPA, LPA' and LPA' with $\omega_k$ in Fig \ref{nu-n=6}. As  $\alpha$ increases in the interval $1<\alpha<5$,  $\nu$ increases  and does not have an extremum. To see whether the absence of an optimal value is due to the particular choice of the parameters, we have calculated $\nu(d=3, N=6, \beta, \alpha=2)$ for several values of the prefactor $\beta$ while fixing the exponent $\alpha=2$ (data not shown) and we did not find an optimal value either. On the other hand, in semi-expansion \cite{NPRG semi-expansion}, $\nu(d=3, N=6)$ varies typically between 0.69 and 0.73 at LPA' with $\omega_k$ depending on the parameter in the exponential regulator. As a function of $\alpha'$ in Eq. (\ref{regexp}), the authors have found that there is no  optimal value of the exponent but for large values of $\alpha'$ the variation of $\nu(d=3, N=6, \alpha')$ is rather small. From this observation, they concluded that an estimate of the exponents is $\nu^{\rm semi}(d=3, N=6)=0.695(5)$ and $\eta^{\rm semi}=0.042(2)$. 
Using the regulator in Eq.~\eqref{reglitim} with $\beta=1$, we find at LPA' with $\omega_k$ a PMS value for $\eta$ at $\alpha\simeq2$: $\eta(\alpha=2)=0.0457$, see Fig.~\ref{etad=3}. For this value of $\alpha$, we find $\nu(\alpha\simeq 2)\simeq0.727$, see Fig.~\ref{nu-n=6}. Provided that the previous results obtained with the semi-expansion were converged, the difference between the two sets of results given above must come from the choice of regulators. We conclude that contrary to the value of $N_c(d=3)$, the variations of the critical exponents with the regulator is not very small, at least for $N=6$. On the other hand, the values of $\nu$ and $\eta$ given above are compatible with the bounds given by the conformal bootstrap \cite{Stergiou}. We have checked that for $N=7$ and $N=8$, our results for both $\nu$ and $\eta$ differ at most by 10$\%$ from the 6-loop $\epsilon$-expansion results and from the Monte Carlo simulations \cite{Kompaniets,Sorokin}. Notice that such an error of about 10$\%$ in $d=3$ is expected at the the level of the LPA' since this is typically what is found in O($N$) models and in models whose upper critical dimension is four \cite{Canet-LPA}.

\subsection{Calculation of $N_c(d=2.5)$}

\begin{figure}[t!]
	\centering 
    \includegraphics[width=0.6\textwidth]{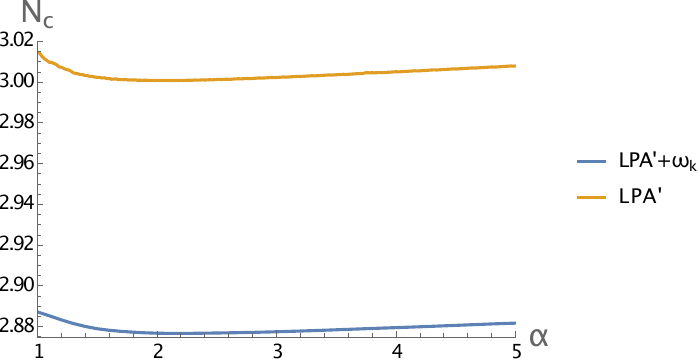}
    \caption{The curve $N_c(d=2.5, \beta=1, \alpha)$ at LPA' and LPA' with $\omega_k$. Here we fix the parameter $\beta=1$  and vary the parameter $\alpha$ in the regulator $R_k(q^2)$. 
    }
    \label{Nc25}
\end{figure}
\begin{figure}[t!]
	\centering 
    \includegraphics[width=0.65\textwidth]{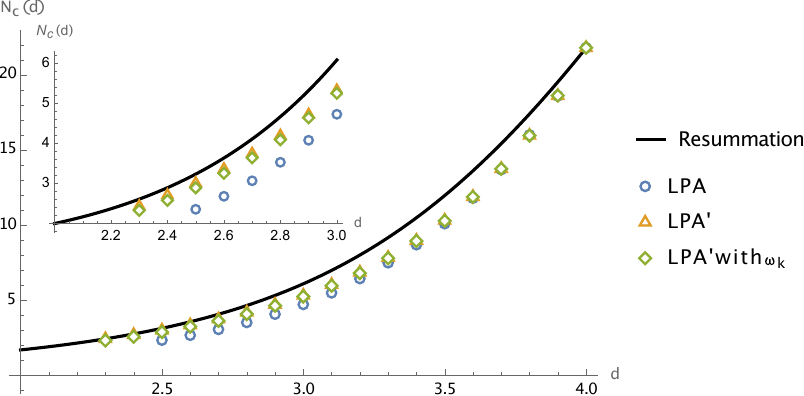}
    \caption{The curve $N_c(d, \alpha=1,\beta=1)$. The continuous curve corresponds to the resummed five-loop $\epsilon$-expansion \cite{Calabrese five-loop}. The spreading of values of $N_c(d, \alpha,\beta)$ when $\alpha$ is varied between 1 and 5 and $\beta$ between 1 and 2 is almost invisible on this scale.}
    \label{Nc}
\end{figure}

As mentioned in the Introduction, it is interesting to study the shape of the curve $N_c(d)$ below $d=3$ to find out whether or not it shows a S-shape and thus a possible re-entrance of the first order region below the value of $N_c(d=3)$ found above. On the example of the dimension $d=2.5$, we show in this section that our solution of the LPA' with $\omega_k$ FP equations is numerically well under control.

As previously emphasized, the LPA is not sufficient to recover the one-loop result in $d=2+\epsilon$ and the LPA' with $\omega_k$ is mandatory. In the following, we focus on this approximation as well as on the LPA' when we study $N_c(d)$ in low dimensions.

We show $N_c(d=2.5, \beta=1, \alpha)$ at LPA' with $\omega_k$ in Fig \ref{Nc25}.   At LPA' and LPA' with $\omega_k$, a minimum as a function of $\alpha$ exists respectively at $N_c^{\rm opt}(d=2.5, \beta=1, \alpha\simeq2)=3.00$ and $N_c^{\rm opt}(d=2.5, \beta=1, \alpha\simeq2)=2.88$. 

\subsection{The curve $N_c(d)$ in $2.3\leq d\leq 4$}

We have found that it becomes more and more difficult to obtain converged results when decreasing the dimension. We have been able to decrease $d$ down to 2.3 but not below with the LPA' with $\omega_k$. \textcolor{black}{As discussed in Appendix B 4, the numerical difficulty is due to the steep increase of the FP potential at large fields. It would be interesting to improve our numerical scheme explained in detail in Appendix B, in the future, in order to approach $d=2$ further.}
 We show our 
 determination of $N_c(d)$  in Fig. \ref{Nc} together with the results obtained from the $\epsilon$-expansion
 at five loops \footnote{The resummation plotted here is done without assuming $N_c(d=2)=2$}. 
 
For $d=3$, our results confirm the previous results obtained either by  NPRG \cite{Zumbach1, Zumbach2, NPRG field expansion, Delamotte Review, NPRG semi-expansion} 
or the $\epsilon$-expansion approaches \cite{Jones, Bailin, Calabrese five-loop}. 
The comparison between the LPA and LPA' results
strongly suggests that neglecting the effect of the derivative terms on the determination of $N_c(d)$ plays
a minor role in $d=3$ as for the order of the transition for $N=2$ and $N=3$. Moreover, $N_c(d=3)$ increases
between the LPA and the LPA' with $\omega_k$ and becomes closer to the results obtained with the $\epsilon$-expansion, which is
expected. It seems  therefore very difficult to imagine that $N_c(d=3)$ could be smaller than 3. Judging from the curve $N_c(d)$ for $2.3\leq d\leq 4$, it also seems very unlikely that $N_c(d)$ has a turn around point and is a multi-valued function around $d=3$, which is one of the expected scenarios to allow for the existence of $C_{+}$ in $d=3$ and $N=3$.   

\subsection{Commments on estimation of the error bars}
\textcolor{black}{Although, throughout this paper, we qualitatively estimate the  error of our calculation using the dispersion of the results when the regulator and the approximation (LPA, LPA' or LPA' with $\omega_k$) are varied, the quantitatively  reliable error bars can only be computed by comparing two successive orders of the derivative expansion. Our calculations, limited to LPA and LPA', are not sufficient to quantitatively estimate the  error bars  of $N_c(d)$. However, the same thing happens for critical exponents with LPA', which nevertheless gives fairly accurate results that are, say, within 10$\%$ of the exact values (as given by the conformal bootstrap).}

\section{Conclusion}
Let us also emphasize that the only Monte Carlo simulations that still find a second order transition for a value of
$N$ below our value of $N_c(d=3)$, that is, for $N\le4$, has been performed for $N=2$ by Calabrese et al. \cite{Calabrese 3} on a discretization
of the Ginzburg-Landau model Eqs.~(\ref{eq:effective hamiltonian}), (\ref{pot}) and by Nagano and Kawamura  \cite{Nagano Kawamura} on Heisenberg STA. Calabrese et al found that depending on the values of the parameters  $\lambda$ and $\mu$ in Eq.~\eqref{pot},
the transition is of first or second order: At fixed $\lambda$ and small $\mu$, the transition is of second order whereas it is of first order
at large $\mu$. Since nonuniversal quantities, such as  phase diagrams \cite{machado,Canet04}, can be accurately 
computed from the integration of the NPRG flow equations, 
it is possible to estimate the magnitude of the correlation length $\xi$ at the  transition within the LPA' 
by initializing the flow with the data corresponding to the simulations. By varying these data as well as the cut-off function
$R_k(q)$, it is found that $\xi$ is always  finite  (since there is no fixed point) but very large, typically larger
than 2000 lattice spacings \cite{Debelhoir}. From a numerical point of view, there is no doubt that  such 
a large correlation length makes it impossible to decide in favor of a second or a very weak first order phase transition since in both cases the physics 
 looks the same at the scale of the lattice size which was at most 384 lattice spacings  in the numerical simulations. 
We conclude that this Monte Carlo result does not
contradict our conclusion that $N_c(d=3)\gtrsim 5$.

This result shows unambiguously that if our result is wrong, the origin of the problem can only be
found by including the renormalization of the functions in front of the derivative terms. However,
considering that the 
anomalous dimension is small for these systems when they undergo a second order phase transition, that is,
for $N>N_c$, this
hypothesis seems very doubtful. 
We suggest that the Blaizot-Mendez-Wschebor approach \cite{BMW1,BMW2,BMW3}, where the full momentum dependence of the 
two-point functions is retained as well as the full field-dependence of the potential $\tilde{U}$ could lead
to a very accurate determination of $N_c(3)$.

As for the approach to $d=2$, we find a remarkable agreement between our results and what was found within
the $\epsilon$-expansion where two resummations  were performed in \cite{Calabrese five-loop},
either by assuming that $N_c(d=2)=2 $ or by letting free
the value of $N_c(d=2) $, see Fig.~\ref{Nc}. This agreement is not very surprising because we expect  the LPA' with $\omega_k$ to be
one-loop exact in $d=2+\epsilon$ for $N>2$. Notice that our results are not precise enough to determine unambiguously
the value of $N_c(d=2) $
although it seems clear that it cannot be very different from 2.
It is therefore very unlikely that $N_c(d=2)>3$. Since the NPRG flow reproduces the low-temperature
expansion of the nonlinear sigma model around $d=2$, we conclude that the critical behavior of frustrated
systems in $d=2+\epsilon$ is driven for $N=3$ by the fixed point $C_+$ corresponding to a critical temperature
of order $\epsilon$ in agreement with Mermin-Wagner theorem. Since we find no other once-unstable
fixed point, we conclude that our study rules out the possibility of having a finite temperature
fixed point in $d=2$ for $N=3$ contrary to what was found at five loops in a fixed
dimension RG calculation \cite{Calabrese focus fixed points 2}.

To conclude, we have presented a rather simple method to compute the FP properties 
of matrix models describing frustrated systems without having recourse to a field expansion
of the free energy $\Gamma$ (but keeping a derivative expansion of $\Gamma$).  This is especially
important in low dimensions where the field expansion is known to fail. In dimension $d=3$, our results
fully confirm what was previously found within less accurate NPRG calculations that involved 
 field truncations on top of the derivative expansion \cite{NPRG field expansion, NPRG semi-expansion, Delamotte Review}. In dimension $d=2$, more stable numerical
 schemes are still needed to study the physics of topological excitations in frustrated systems
 (that are of different natures than in nonfrustrated systems) and we believe that the present work
 is the first step in this direction. With field expansion, we studied the fate of the $C_{-}$ FP in dimensions lower than 3 and found that it vanishes by colliding with another muticritical FP \cite{ON2017}. Since multicritical FPs show in their FP potential  boundary layer at Large-$N$ in $O(N)$ models \cite{ON2018, ON2022}, which needs functional treatment, it would be also important to study functionally the multicritical FPs in the frustrated cases. 
 Finally, let us mention that there are many systems with multiple invariants whose two-dimensional physics can only be studied functionally in the NPRG framework \cite{Chlebicki}. Similar procedures to that presented in this work might be useful for such models.

\section{Acknowledgment}

This work was supported in part by a Grant-in-Aid for Young
 Scientists
(B) (15K17737), Grants-in-Aid for Japan Society for Promotion of Science (JSPS) Fellows (Grants Nos. 241799 and 263111), the JSPS Core-to-Core Program "Non-equilibrium dynamics of soft matter and information".

\appendix

\section{The nonperturbartive renormalization group flow equations and the anomalous dimension}

Since the explicit expression of flow equations  is rather complicated except when the Litim regulator ($\alpha=\beta=1$) is taken and $\omega_k=0$, we present them only for $\alpha=\beta=1$ and $\omega_k=0$.  See \cite{NPRG semi-expansion} for the complete expressions  at LPA' with $\omega_k\ne0$.
Then, the running anomalous dimension $\eta_{k}=-k\partial_{k}Z_{k}$
is given, at the level of LPA', by

\begin{eqnarray}
\eta_{k} & = & 64\frac{\tilde{\kappa}v_{d}}{d}\left(2\left(\frac{\tilde{U}_{k}^{\left(2,0\right)'}}{1+4\tilde{\kappa}\tilde{U}^{\left(2,0\right)'}}\right)^{2}+\left(\frac{\tilde{U}_{k}^{\left(0,1\right)'}}{1+2\tilde{\kappa}\tilde{U}_{k}^{\left(0,1\right)'}}\right)^{2}\right),\nonumber \\
\label{eq:eta}
\end{eqnarray}
where we set $\tilde{\rho}=\tilde{\kappa}$ and $\tilde{\tau}=0$.
The derivatives $\tilde{U}_{k}^{\left(i,j\right)'}$ with respect
to the invariants $\tilde{\rho}$ and $\tilde{\tau}$, and $v_{d}$
are defined as 
\[
\tilde{U}_{k}^{\left(i,j\right)'}\equiv\frac{\partial^{i+j}\tilde{U}_{k}}{\partial\tilde{\rho}^{i}\partial\tilde{\tau}^{j}},\ v_{d}=\frac{1}{2^{d+1}\pi^{d/2}\Gamma\left(\frac{d}{2}\right)}.
\]

The nonperturbartive renormalization group flow equation for the potential $\tilde{U}_{k}$ is
given by
\begin{eqnarray}
\partial_{t}\tilde{U}_{k} & = & -d\tilde{U}_{k}+\frac{1}{2}(-2+d+\eta_{k})\left(\tilde{\psi}_{1}\tilde{U}_{k}^{\left(1,0\right)}+\tilde{\psi}_{2}\tilde{U}_{k}^{\left(0,1\right)}\right)\nonumber \\
 &  & +\frac{4(2+d-\eta_{k})}{d(2+d)}v_{d}\nonumber \\
 &  & \times\left(\frac{\tilde{\psi}_{1}-\tilde{\psi}_{2}}{\tilde{\psi}_{1}-\tilde{\psi}_{2}-\tilde{U}_{k}^{\left(0,1\right)}+\tilde{U}_{k}^{\left(1,0\right)}}+\frac{\tilde{\psi}_{1}+\tilde{\psi}_{2}}{\tilde{\psi}_{1}+\tilde{\psi}_{2}+\tilde{U_{k}}^{\left(0,1\right)}+\tilde{U}_{k}^{\left(1,0\right)}}\right.\nonumber \\
 &  & +(N-2)\left(\frac{\tilde{\psi}_{2}}{\tilde{\psi}_{2}+\tilde{U}_{k}^{\left(0,1\right)}}+\frac{\tilde{\psi}_{1}}{\tilde{\psi}_{1}+\tilde{U}_{k}^{\left(1,0\right)}}\right)\nonumber \\
 &  & \left.+{\normalcolor }\frac{2+\tilde{U_{k}}^{\left(0,2\right)}+\tilde{U}_{k}^{\left(2,0\right)}}{1-\left(\tilde{U}_{k}^{\left(1,1\right)}\right)^{2}+\tilde{U}_{k}^{\left(2,0\right)}+\tilde{U}_{k}^{\left(0,2\right)}\left(1+\tilde{U}_{k}^{\left(2,0\right)}\right)}\right).\label{eq:flow eq}
\end{eqnarray}
Here, to simplify the notation, we have defined another kind of derivatives $\tilde{U}_{k}^{\left(i,j\right)}$ with respect
to $\tilde{\psi}_{1}$ and $\tilde{\psi}_{2}$ as
\begin{equation}
\tilde{U}_{k}^{\left(i,j\right)}\equiv\frac{\partial^{i+j}\tilde{U}_{k}}{\partial\tilde{\psi}_{1}^{i}\partial\tilde{\psi}_{2}^{j}}.
\label{eq:flowU}
\end{equation}

In our calculations, we use the rescaled potential ${v_{d}^{-1}\tilde{U}_{k}}$ and fields $\left(v_{d}\right)^{-1/2}\tilde{\psi}_{i}$
for $i=1,2$ 
in such a way that $v_{d}$ disappears in Eqs (\ref{eq:eta}) and
(\ref{eq:flow eq}).

\section{Numerical methods}

\subsection{The fixed point at LPA}\label{numerical-scheme} 

To simplify the explanation, we focus on LPA in this subsection. From a numerical point of view, there are two possibilities for finding
fixed points when they exist. 

The first is to dynamically integrate the flow. In this
case, the problem is to find the critical surface which is usually
done by dichotomy on the temperature. Once it is found, the fixed point is (approximately)
reached since it is attractive on the critical surface. The approximation of the FP found this way can be used as an initial condition for a direct search of the FP by a Newton-Raphson method if it is necessary to refine the numerical  accuracy. 

The other
method is to look directly for the solution of the fixed point equation
(coupled with Eq. (\ref{eq:eta})): 
$\partial_{t}\tilde{U}^{*}(\tilde{\psi}_{i})=0$.
This is what we do here.  The advantage of this method is three-fold:
(i) The numerical scheme is much simpler than integrating the flow; 
(ii) several numerical instabilities
occuring during the integration of the flow are avoided; (iii) the
critical exponents are easily obtained from the diagonalization of
the RG flow around the fixed point. We show in the following that 
although this scheme works very well in dimension $d=3$, numerical difficulties arise in dimensions close to $d=2$, making it almost impossible to study the physics of frustrated systems in this dimension, at least with our numerical scheme.

The basic idea of this scheme is simple. It consists in solving the FP equations for $\tilde{U}^{*}$
on a grid in $(\psi_{1},\psi_{2})$ space, taking into account
the symmetries (\ref{constraint}). We introduce a cut-off field value $\tilde{\psi}_{max}$
and consider the triangular domain $D:\tilde{\psi}_{max}\geq\tilde{\psi}_{2}\geq\tilde{\psi}_{1}\geq0$.
We then discretize $D$ on a square lattice with mesh size $\Delta\tilde{\psi}=\tilde{\psi}_{max}/\left(N_{p}-1\right)$,
where $N_{p}$ is the number of lattice points on the axis $\psi_{1}=0$.
The lattice points are given by $\left(i\Delta\tilde{\psi},j\Delta\tilde{\psi}\right)$
for integers $i$ and $j$ that satisfy $0\leq i\leq j\leq N_{p}-1$.
We define $\tilde{U}_{t}\left(i,j\right)\equiv\tilde{U}_{t}\left(i\Delta\psi,j\Delta\psi\right)$
to alleviate the notation.

The fixed point equation for the potential is a differential equation. We transform
it into a set of algebraic equations by discretizing the derivatives of $\tilde{U}$.
We give below some details about this procedure because all our numerical
problems come from the boundary of the domain $D$, precisely at the
points where the discretization involves exceptional cases.

The formulae for the derivatives $\tilde{U}_{t}^{\left(l,m\right)}\left(i,j\right)$
for $l,m=0,1,2$ are constructed as follows:

(1) In the bulk region ($0\leq i\leq j\leq N_{p}-3$): $U^{\left(1,0\right)}$
and $U^{\left(2,0\right)}$ as well as $U^{\left(0,1\right)}$ and
$U^{\left(0,2\right)}$ are computed  with  five points.
$U^{\left(1,1\right)}$ is computed with the nine points $\tilde{U}_{t}\left(i,j\right)$,
$\tilde{U}_{t}\left(\left(i\pm1\right),\left(j\pm1\right)\right)$,
$\tilde{U}_{t}\left(\left(i\pm1\right),\left(j\mp1\right)\right)$,
$\tilde{U}_{t}\left(\left(i\pm2\right),\left(j\pm2\right)\right)$
and $\tilde{U}_{t}\left(\left(i\pm2\right),\left(j\mp2\right)\right)$.
The formulae are exact up to $\left(\Delta\psi\right)^{3}$.
Notice that for points on the two borders of $D$ defined either
by $\tilde{\psi}_{1}=0$ or $\tilde{\psi}_{1}=\tilde{\psi}_{2}$,
the derivatives of $\tilde U$ involve points outside $D$. By using (\ref{constraint}),
we can compute these values of $\tilde U$ from those that are inside $D$. This is one 
of the advantage of the choice of variables $(\psi_{1},\psi_{2})$ compared
to the choice $(\rho, \tau)$: The derivatives on the two borders 
$\tilde{\psi}_{1}=0$ and $\tilde{\psi}_{1}=\tilde{\psi}_{2}$ can be computed
in the same way as in the bulk.

(2) On the boundary of the domain $D$ corresponding to the large
field region, $j=N_{p}-2,N_{p}-1$, we compute the derivatives in the $\psi_1$ direction
$U^{\left(1,0\right)}(i,j)$
and $U^{\left(2,0\right)}(i,j)$ in the same way as in (1), that is, as in the bulk. The formulae
for $U^{\left(0,1\right)}(i,j)$ and $U^{\left(0,2\right)}(i,j)$
are constructed with the five quantities $\tilde{U}_{t}\left(i,j'\right)$
for $j'=N_{p}-5,\cdots,N_{p}-1$ and are exact at order $\left(\Delta\psi\right)^{2}$.
The formula for $U^{\left(1,1\right)}\left(i,N_{p}-1\right)$ for
$0\leq i\leq N_{p}-2$ involves the six values $\tilde{U}_{t}\left(i+1,j'\right)$,
$\tilde{U}_{t}\left(i-1,j'\right)$ for $j'=N_{p}-3,N_{p}-2,N_{p}-1$
and is exact at order $\left(\Delta\psi\right)$. Finally, for $U^{\left(1,1\right)}\left(N_{p}-1,N_{p}-1\right)$
we use twelve points in the region $N_{p}-4\leq i\leq j\leq N_{p}-1$
and the formula is exact at order $\left(\Delta\psi\right)^{2}$.

Notice that we have increased the precision
of the derivatives on the boundary of the domain $D$ corresponding
to the large field region in order to test the robustness of our results with respect
to the choice of discretization and to try to reduce numerical problems when $d$ is close
to 2. In all cases studied we did not find any significant changes. In particular,
the scheme is not more stable when the number of points chosen to compute the derivatives is increased.

Once the derivatives are discretized, the fixed point equation
$\partial_{t}\tilde{U}^{*}\left(\psi_{1},\psi_{2}\right)=0$
becomes a set of coupled algebraic equations for $g_{i,j}^{*}\equiv\tilde{U}^{*}\left(i,j\right)$.
We look for a solution to these equations by a Newton-like method. One of the difficulty of this method
is the huge number of unknowns and the possibility for Newton's method
to get lost in the very complicated landscape of extrema of the set
of equations to be solved. The way out of this difficulty is to deform continuously
a solution of the problem.

\begin{figure}[t!]
	\centering 
    \includegraphics[width=0.45\textwidth]{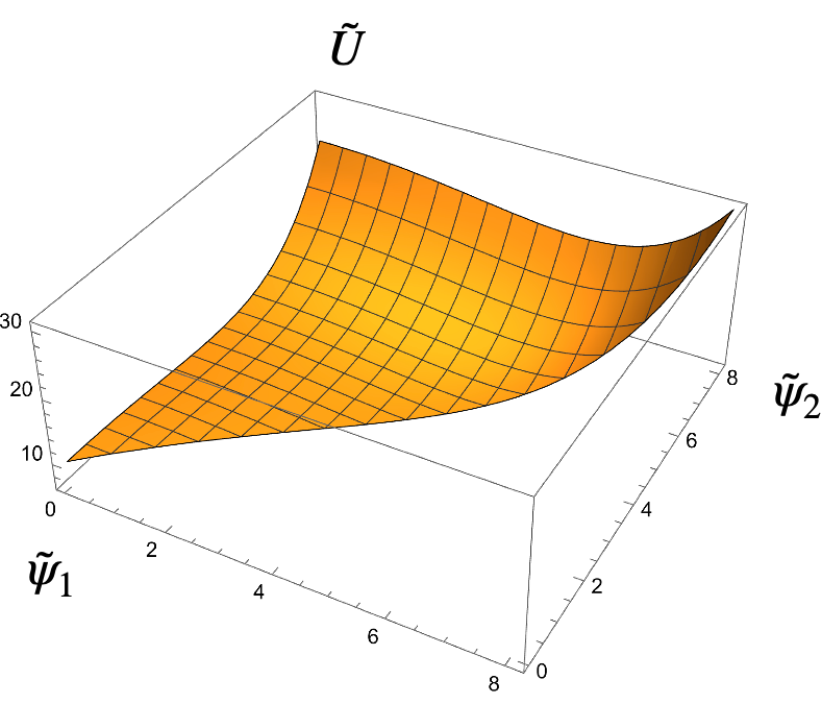}

    \includegraphics[width=0.45\textwidth]{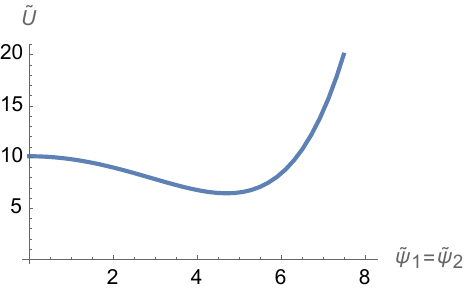}
    \caption{(Top) The FP potential $\tilde{U}^*(\tilde{\psi}_1,\tilde{\psi}_2)$ of the $C_{+}$ FP for $d=3$ and $N=8$ at LPA. (Bottom) The FP potential $\tilde{U}^*(\tilde{\psi}_1,\tilde{\psi}_1)$ on the diagonal line $\tilde{\psi}_1=\tilde{\psi}_2$ of the same FP. The minimum of the potential locates at $(\tilde{\psi}_1,\tilde{\psi}_2)=(4.68,\,4.68)$.}
    \label{FPpot}
\end{figure}

Our strategy in this paper is to follow the fixed point potential
$\tilde{U}^{*}\left(\tilde{\psi}_{1},\tilde{\psi}_{2}\right)$ by changing
the dimension $d$ and the number of spin components $N$ gradually
starting from $d=3.9$ and $N=22$ where the
field-expansion method provides a good approximation of the fixed
point potential. We use as an initial condition of  Newton's method: 
\begin{equation}
\tilde{U}^{*,\,{\rm init}}\left(\tilde{\psi}_{1},\tilde{\psi}_{2}\right)=
\frac{\tilde{\lambda}^{*}}{2}\left(\tilde{\rho}-\tilde{\kappa}^{*}\right)^{2}+
\tilde{\mu}^{*}\tilde{\tau}
\label{eq:field expansion}
\end{equation}
and $\eta=0$.
The parameters  $\tilde{\lambda}^{*}$, $\tilde{\kappa}^{*}$ and $\tilde{\mu}^{*}$
are  determined by performing a field-expansion of the LPA equation on $\tilde{U}$ 
at order four in the fields and solving the fixed point equation 
for these parameters in $d=3.9$ and for $N=22$. As expected, we find four fixed points: the Gaussian and 
the $O(2N)$ fixed points as well as a once-unstable fixed point $C_{+}$ driving the phase transition
and $C_{-}$ that corresponds to a tricritical fixed point. Once an approximation of $C_{+}$ is found 
with the truncation of Eq. (\ref{eq:field expansion}), we use it as the initial condition of Newton's method
for the full potential equation 
and we easily find $\tilde{U}^*$.
Then, we move in the $(d,N)$ plane by little steps using as new initial condition what was found
for the previous value of $d$ and/or $N$ studied. The fixed potential potential deforms smoothly and 
the Newton's method always works properly this way. We give a plot of the FP potential for the $C_+$ FP for $d=3$ and $N=8$ at LPA in Fig. \ref{FPpot}.

\subsection{Eigenvalues of the stability matrix around a fixed point}

For each value of $(d,N)$ studied, we  compute the eigenvalues of the stability
matrix $\Theta\left(\left\{ i,j\right\} ,\left\{ i',j'\right\} \right)$
defined as 
\begin{equation}
\Theta\left(\left\{ i,j\right\} ,\left\{ i',j'\right\} \right)
\equiv\frac{\partial\left(\partial_{t}g_{\left\{ i,j\right\} }\left(t\right)\right)}
{\partial g_{\left\{ i',j'\right\} }\left(t\right)}|_{g_{i,j}^{*}}
\end{equation}
 where we consider $\left\{ i,j\right\} $ and $\left\{ i',j'\right\} $
as (super-)indices. Since the RG time $t=\log k/\Lambda$ is negative,
a negative (positive) eigenvalue of the matrix $\Theta$ corresponds
to a relevant (irrelevant) eigendirection around the fixed point. We sort
the eigenvalues as $\sigma_{0}\left(=-d\right)<\sigma_{1}<\cdots<\sigma_{i-1}<\sigma_{i}<\cdots$.
Note that the above stability matrix around any fixed point solution
has a trivial relevant eigendirection corresponding to the constant shift $g_{i,j}=g_{i,j}^{*}+\mathrm{const}$
with the eigenvalue $\sigma_{0}=-d$, which can be easily seen from
Eq. (\ref{eq:flowU}). Hereafter, this trivial eigenvalue
is omitted when we discuss the stability of a fixed point. The critical
exponent $\nu$ is given by $\nu=-1/\sigma_{1}$ and the smallest positive 
eigenvalue that vanishes when $C_+$ collapses with $C_-$ is $\sigma_{2}$.

\subsection{Extension to LPA' and LPA' with $\omega_k$}

In this subsection we briefly explain how we extend, for LPA' and LPA' with $\omega_k$, the procedure to calculate the FP solution and the eigenvalues of the stability matrix at LPA. At LPA' and LPA' with $\omega_k$, we have to take into account $\eta\ne0$ and the flow of $\omega_k$, which are evaluated at the minimum of the potential $(\rho,\tau)=( \kappa, 0)$. Note that $\kappa$ is determined as a function of $g_{i,j}$. Therefore the coupling constants that we have to consider here are $g_{i,j}$ and $\omega_k$. The flow of $g_{i,j}$ or $\omega_k$ can be obtained as a function of $g_{i,j}$ or $\omega_k$ with the same discretization procedure described so far. 
At a given $(d,N)$, by recursively adjusting $\tilde{\psi}_{max}$, the minimum of the potential locates exactly on a lattice point, so that all the derivatives $U^{(k,l)}_{i,j}$ at the minimum can be evaluated with the procedure explained in the subsection \ref{numerical-scheme}. With this $\tilde{\psi}_{max}$, we find the FP solution. To evaluate the stability matrix for these flow equations, we  calculate $\frac{\partial \kappa \left(t\right)}{\partial\left(\partial_{t}g_{\left\{ i,j\right\} }\left(t\right)\right)}
|_{g_{i,j}^{*},\omega_k}$ as follows: The perturbation around the FP $g_{i,j} \rightarrow g_{i,j}^*+\delta  g_{i,j}$ changes the location of the minimum as $\kappa \rightarrow \kappa+\delta \kappa$. Since ${\partial (U_k^*+ \delta U_k) (\rho=\kappa+\delta \kappa, 0)}/{\partial \rho}=0$, at the linear order of perturbation, we have
\begin{equation}
\delta\kappa=-\frac{\partial \delta U\left(\rho=\kappa,0\right)}{\partial\rho}/\frac{\partial^{2}U^{*}\left(\rho=\kappa,0\right)}{\partial\rho^{2}},
\end{equation} which can be evaluated in terms of $g^*_{i,j}$ and $\delta g_{i,j}$ using the expressions of the discretized derivatives of $U$ presented in the previous subsection \ref{numerical-scheme}. This leads to the desired discretized expression of $\frac{\partial \kappa \left(t\right)}{\partial\left(\partial_{t}g_{\left\{ i,j\right\} }\left(t\right)\right)}
|_{g_{i,j}^{*},\omega_k}$. 

\subsection{Numerical instabilities}
\begin{table}
\begin{equation}
\begin{array}{c|c}
N_p=61  &   -3, -1.45, 0.218, 0.827, 1.99, 2.79\\
        &      -0.464\pm 34.8i, 0.250\pm 30.9i, 1.07\pm 27.6i\\
N_p=81  &   -3 ,-1.45, 0.218, 0.827, 1.99, 2.79\\
        &       0.059 \pm 47.9i, 0.868 \pm 43.5i, 1.76 \pm 39.9 i\\
N_p=101 &   -3, -1.45, 0.218, 0.827, 1.99, 2.79,\\
        &       0.704\pm 61.05i,1.627\pm 56.3i\\
\end{array}
\end{equation}
\caption{  Several of the most relevant eigenvalues around
the $C_{+}$ fixed point for $N=5$
and $d=3$ at LPA with Litim cutoff. We have chosen $\tilde{\psi}_{max}=9/4\tilde{\psi}_{min}$.
 The physical eigenvalues are given on the first line for each value of $N_p$
and the others, that are spurious, on the second line. For $N_p=61$, the eigenvalues  $-0.464\pm 34.8i$
are relevant since their real part is negative. This eigenvalue disappears when increasing $N_p$, as it should.}
\label{eigenvalues} 
\end{table}
For each dimension $d$ and value of $N$ we have to make sure that our results are converged.
First we focus on LPA with Litim cutoff. Once the choice of discretization of the derivatives has been made, there are two parameters that
can be tuned: the values of $\tilde{\psi}_{max}$ and of the mesh size
$\Delta\tilde{\psi}=\tilde{\psi}_{max}/\left(N_{p}-1\right)$. The potential $\tilde{U}^*$ shows
a minimum at $\tilde{\psi}_1=\tilde{\psi}_2=\tilde{\psi}_{min}$ and we have observed that $\tilde{\psi}_{max}$
should be at least 1.5 times larger than $\tilde{\psi}_{min}$ to get values of $N_c(d)$ converged with an accuracy
of at least $1\%$. We have also observed that the smaller  the 
dimension, the smaller  $\Delta\tilde{\psi}$ must be to get converged results. 
This last point has two origins. First, for $d$ close to 2, the FP potential is steep at large fields because
it behaves as $\left(\tilde{\psi}_{1}^{2}+\tilde{\psi}_{2}^{2}\right)^{\frac{d}{d-2+\eta}}$ and a small mesh size
is necessary to accurately describe the shape of $\tilde{U}^*$. Second, if $N_p$ is too small, 
we find that even far away from $d=2$,
say $d=3$,  several eigenvalues corresponding to relevant eigendirections appear in the spectrum and
spoil the degree of stability of the fixed point $C_+$. These eigenvalues
are clearly spurious because their values change considerably when either $\Delta\tilde{\psi}$
is decreased or $\tilde{\psi}_{max} $ is increased whereas the complementary set
of eigenvalues, the physical ones,  remain unchanged up to the sixth digit, see Table \ref{eigenvalues}.
We observe that 
as $\Delta\tilde{\psi}$ is decreased, these spurious eigenvalues systematically disappear (or, at least, get
a very large real part which makes them highly irrelevant). 
The conclusion of this study is that
for each $d$, a sufficiently large $N_p$ should be chosen so that the set of first most relevant eigenvalues
is converged as for their numbers and values. In particular, when the numerical results are converged all spurious relevant eigenvalues have disappeared. We find that in $d=3$, $N_p=101$ is sufficient to get fully
converged results while leading to numerically feasible calculations, when we take $\tilde{\psi}_{max}=2.27 \tilde{\psi}_{min}$. We also find that as $d$ approaches 2,  
``large'' values
of $\tilde{\psi}_{max}$ favor the presence of spurious eigenvalues that can only be eliminated by
increasing $N_p$. It turns out  that around $d=2.4$, very large values of $N_p$, such as $N_p=200$,
would be necessary to avoid spurious eigenvalues and that decreasing $d$ would impose to increase $N_p$
in a prohibitive way. This is why we did not determine $N_c(d) $ at LPA for $d<2.5$.

Next we discuss LPA' and LPA' with $\omega_k$. Note that the problem of spurious eigenvalues encountered at LPA becomes less severe for LPA' and LPA' with $\omega_k$, since $\eta\ne 0$ which implies that the FP potential is less steep at large fields than at LPA. Typically, when we take $\tilde{\psi}_{max}\simeq1.5 \tilde{\psi}_{min}$, $N_p=40$ in $d=2.4$ is sufficient to avoid spurious eigenvalues with negative real part. We have been able to compute $N_c(d)$ down to $d=2.3$ by computing directly
the value of $N$ where no fixed point $C_+$ is found with Newton's method
but we have not been able to go below this dimension.

\end{document}